\documentclass[aps,showpacs,superscriptaddress,preprint]{revtex4}%
\usepackage{amsfonts}
\usepackage{amsmath}
\usepackage{amssymb}
\usepackage{graphicx}%
\begin{document}

\title{The equation of state and nonmetal-metal transition of benzene under shock compression}

\author{Cong Wang}
\affiliation{LCP, Institute of Applied Physics and Computational
Mathematics, P.O. Box 8009, Beijing 100088, People's Republic of
China}
\author{Ping Zhang}
\thanks{Corresponding author. zhang\_ping@iapcm.ac.cn}
\affiliation{LCP,
Institute of Applied Physics and Computational Mathematics, P.O. Box
8009, Beijing 100088, People's Republic of China}
\affiliation{Center for Applied Physics and Technology, Peking
University, Beijing 100871, People's Republic of China}

\begin{abstract}
We employ quantum molecular dynamic simulations to investigate the
behavior of benzene under shock conditions. The principal Hugoniot
derived from the equation of state is determined. We compare our
firs-principles results with available experimental data and provide
predictions of chemical reactions for shocked benzene. The
decomposition of benzene is found under the pressure of 11 GPa. The
nonmetal-metal transition, which is associated with the rapid C-H
bond breaking and the formation of atomic and molecular hydrogen,
occurs under the pressure around 50 GPa. Additionally, optical
properties are also studied.
\end{abstract}

\pacs{62.50.-p, 71.30.+h, 31.15.xv} \maketitle

\section{INTRODUCTION} \label{sec-intr}

The nature of hydrocarbons under extreme conditions is of great
scientific interest and has recently attracted extensive studies.
Specially, due to the possible importance in the formation of
organic materials in the early solar system \cite{PBX:Elert:2003},
chemical reactions of small hydrocarbon molecules, such as
acetylene, ethylene, methane, and benzene have gained particular
attention. A major constitute of the \textquotedblleft
ice\textquotedblright\ layer in Uranus and Neptune, where the
pressure could reach 600 GPa and temperature 8000 K, is thought to
be methane \cite{PBX:Hubbard:1980,PBX:Stevenson:1982}, and the
planetary magnetic fields are greatly influenced by the electrical
conductivity of methane at these conditions. Another important
source of planetary carbon is thought to benzene, which has been
found in the atmosphere of Jupiter and in carbonaceous chondrites
\cite{PBX:Mimura:1994,PBX:Mimura:1995a}. Naturally, comprehensive
studies on the characteristics of benzene are indispensable for
understanding the various astrophysical phenomena at high
temperatures and pressures.

Researches on benzene properties under shock conditions start from
the plate impact experiments \cite{PBX:Dick:1969}. As pressure
increases, benzene has been found to decompose at 13 GPa along its
Hugoniot
\cite{PBX:Hauver:1965,PBX:Dick:1970,PBX:Warnes:1970,PBX:Yakusheva:1971,PBX:Holmes:1990}.
At high static and dynamic temperatures and pressures, benzene
mainly converts into defective carbon nanoparticles and H$_{2}$,
with the possible products of small concentrations of polycyclic
aromatic hydrocarbons, such as alkanes
\cite{PBX:Nellis:1984,PBX:Mimura:1995b}. Such reactions have also
been found to be similar to the detonation of explosive materials
\cite{PBX:Davis:1996}. Nonmetal-metal transition of benzene was
experimentally reported by Nellis \emph{et al.}
\cite{PBX:Nellis:2001} and was attributed to thermal activation of
defective carbon nanoparticles under the pressure between 20 and 40
GPa; further increase in pressure does not have much affect on the
conductivities. To date, although a number of explanatory and
predictive results in some cases have already been provided by
experimental studies, less information is available for
shock-introduced chemical reactions, and many fundamental questions
of benzene under extreme conditions are still yet to be
investigated.

Concerning the theoretical side, a tight-binding model has been
applied to study the properties of liquid benzene
\cite{PBX:Bickham:2000}. The results agree well with the experiments
in ($D$, $u$) diagram (shock velocity verse particle velocity), but
the calculated temperatures remain too high. The thermodynamic
properties of benzene were taken into account by classical force
field method \cite{PBX:Maillet:2008}. While this method showed good
agreement with the low-pressure experimental measurements, however,
it is unsuitable in determining Hugoniot curve at higher pressures.
Overall, due to the intrinsic approximations of these methods, the
electronic structure, which is predominant in determining the
nonmetal-metal transition and equation of state (EOS) for benzene
under shock conditions, is out of consideration. Thus, the change in
the electronic structure of benzene under extreme condition is still
recommended to be presented with a full quantum-mechanical
description. On the other side, quantum molecular dynamics (QMD),
where electrons are fully quantum mechanically treated, has been
proven particularly suitable for the study of the chemical
reactions, such as dissociation, ionization, and recombination of
molecules. Meanwhile, the electronic structure, thermodynamical and
optical properties of warm dense matter have been successfully
investigated by QMD simulations
\cite{PBX:Mattsson:2006,PBX:Kietzmann:2007}. According to the
aspects mentioned above, the electronic structure, EOS and Hugoniot
from density functional theory (DFT) are highly needed for shocked
benzene.

In the present work, we apply QMD simulations to study the
properties of benzene along the principal Hugoniot. The EOS and the
pair correlation functions (PCF) are determined by QMD simulations.
The Kubo-Greenwood formula is used as a starting point for
calculating the dynamic conductivity $\sigma(\omega)$, from which
the dc conductivity is determined. The dielectric function
$\epsilon(\omega)$ and reflectivity are then extracted. This paper
is organized as follows. The simulation details are briefly
described in Sec. \ref{sec-qmd}; The PCF, which is used to study the
dissociation of benzene, and Hugoniot curve are given in Sec.
\ref{sec-hugoniot}; In Sec. \ref{sec-optics} Nonmetal-metal
transition and optical properties are discussed. Finally, we close
our paper with a summary of our main results.

\section{COMPUTATIONAL METHOD} \label{sec-qmd}

In this study, we perform simulations for benzene by employing the
Vienna Ab-initio Simulation Package (VASP) plane-wave
pseudopotential code, which is developed at the Technical University
of Vienna \cite{PBX:Kresse:1993,PBX:Kresse:1996}. Electrons are
fully quantum mechanically treated through plane-wave,
finite-temperature DFT \cite{PBX:Lenosky:2000,PBX:Bagnier:2001}, and
the electronic states are populated according to the Fermi-Dirac
distribution at temperature $T_{e}$. The exchange correlation
functional is determined by generalized gradient approximation (GGA)
with the parametrization of Perdew-Wang 91 \cite{PBX:Perdew:1991}.
The ion-electron interactions are represented by a projector
augmented wave (PAW) pseudopotential
\cite{PBX:Blochl:1994,PBX:Kresse:1999}. Atoms move classically
according to the forces, which originate from the interactions of
ions and electrons. The system is calculated with the isokinetic
ensemble (NVT). In all the simulations, the system is kept in local
thermodynamical equilibrium by setting the electron ($T_{e}$) and
ion ($T_{i}$) temperatures equal. The ion temperature $T_{i}$ is
kept constant every time step by velocity scaling.

For molecular dynamic simulations, only  $\Gamma$ point is
employed to sample the Brillouin zone, while 4$\times$4$\times$4
Monkhorst-Pack \cite{PBX:Monkhorst:1976} scheme $k$ points are
used for the electronic structure calculations. The plane-wave
cutoff energy is set to be 600.0 eV. 48 carbon and 48 hydrogen
atoms (8 benzene molecules) are calculated in a cubic cell at
separate densities and temperatures. The selected densities range
from 1.2 to 2.7 g/cm$^{3}$ and temperatures between 500 K and 6000
K to present the principal Hugoniot. All the dynamic simulations
are lasted for 4$\sim$6 ps, and the time step selected for the
atomic motion is 2 fs. Then, the system is equilibrated 200 steps
and the properties are calculated via the data from the final 300
steps.

\section{SHOCK EQUATION OF STATE AND PAIR CORRELATION FUNCTION} \label{sec-hugoniot}

The accurate calculation of electrical and optical properties
depends on a precise description of the materials properties, such
as EOS. A crucial measurement of EOS data of benzene under shock
condition is the Hugoniot \cite{PBX:Zeldovich:1966}, which can be
derived from conservation of matter, momentum, and energy for an
isolated system compressed by a pusher at a constant velocity.
Rankine-Hugoniot equation describes the locus of points in ($E$,
$P$, $V$)-space satisfying the relation as follows:
\begin{equation} \label{hugoniot}
    (E_{0}-E_{1})+\frac{1}{2}(V_{0}-V_{1})(P_{0}+P_{1})=0
\end{equation}
where $E$ is the internal energy, $V$ is the volume, $P$ is the
pressure, and the subscripts 0 and 1 refer to the initial and
shocked state, respectively. In the canonical ensemble, shock
adiabat between the initial and final states is described by the
principal Hugoniot, which includes the locus of states ($E$, $P$,
$V$) satisfying Eq. (\ref{hugoniot}). In our present calculations,
the pressure $P$ is evaluated using the forces provided by VASP. The
internal energy consists of the total energy from finite-temperature
DFT calculation and zero-point energy. The determination of Hugoniot
points is described as follows. For a given $V_{1}$, a series of
simulations have been executed for different temperatures $T$. Then
$E_{1}$ and $P_{1}$ are fitted to a cubic function of $T$. The
principal Hugoniot points ($E_{1}$, $P_{1}$, $V_{1}$) are evaluated
by solving Eq. (\ref{hugoniot}). For the present system, the initial
density $\rho_{0}$ is 0.874 g/cm$^{3}$, and the internal energy
$E_{0}$= 72.83 eV/molecule at the temperature $T$=298 K. The initial
pressure can be neglected compared to the high pressure of shocked
states. Principal Hugoniot points derived from Eq. (\ref{hugoniot})
are listed in Tab. \ref{H_data}.

\begin{figure}[!htbp]
\centering
\includegraphics*[width=\textwidth]{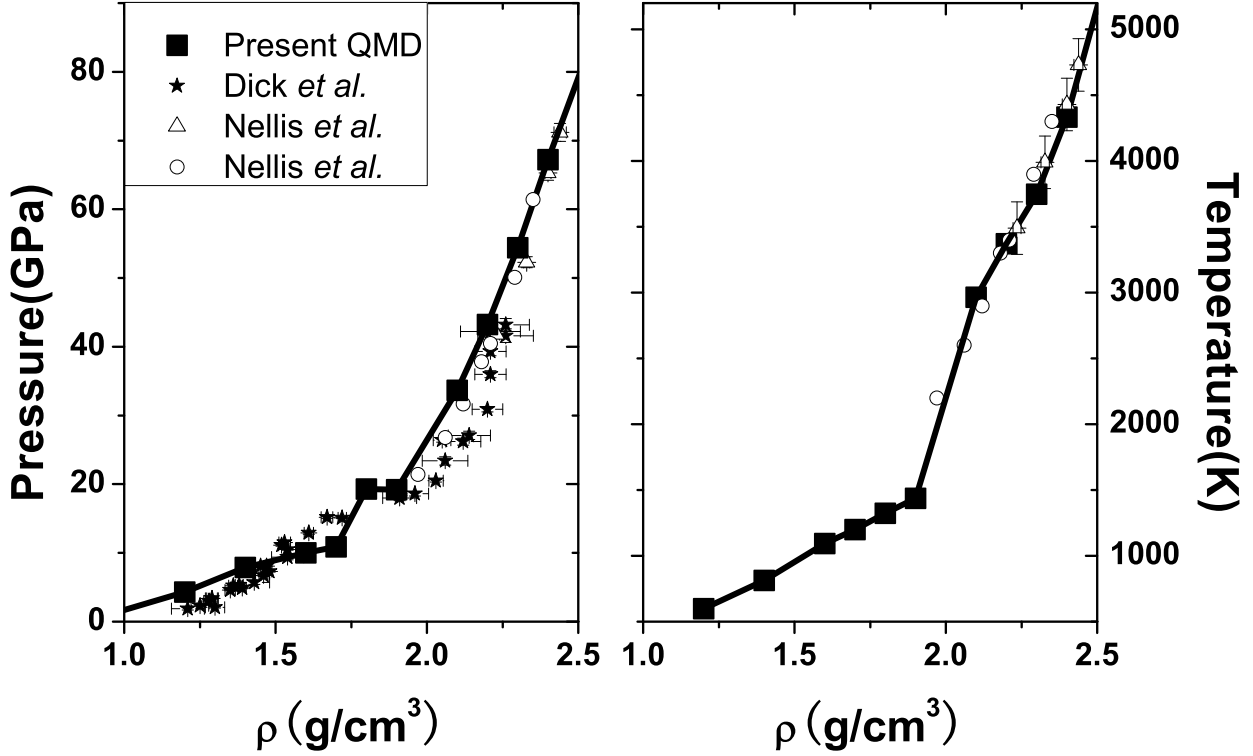}
\caption{Comparison between the previous experimental data of Refs.
\cite{PBX:Dick:1969,PBX:Nellis:1984,PBX:Nellis:2001} and the present
QMD results for the principal Hugoniot.} \label{hugoniot_fig}
\end{figure}

\begin{figure}[!htbp]
\centering
\includegraphics*[width=10 cm]{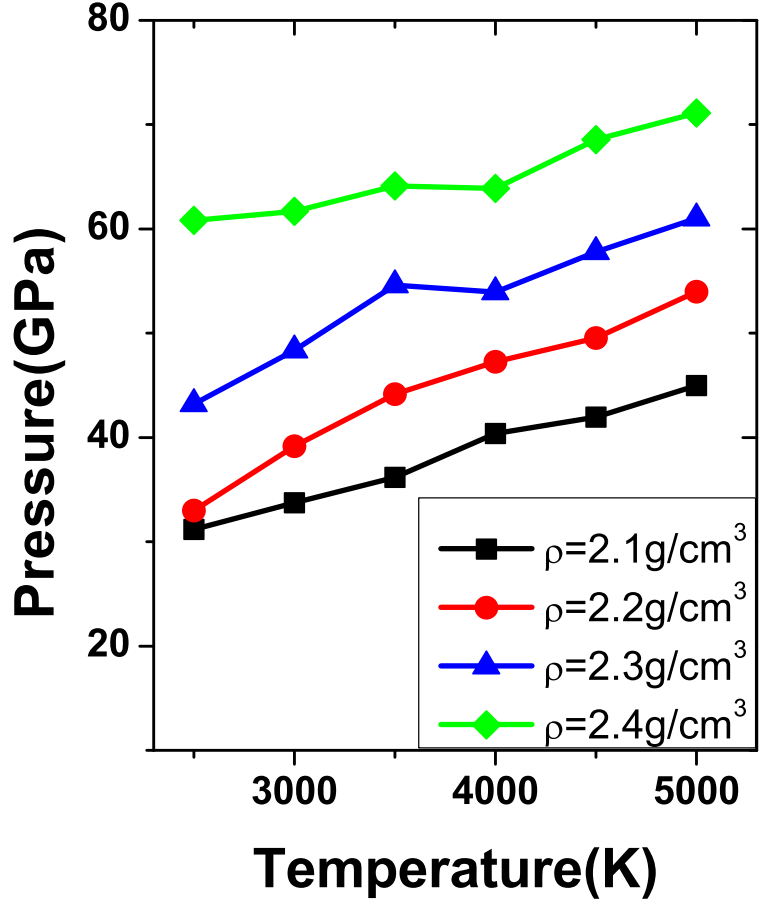}
\caption{$P$-$T$ curves for four densities of benzene.} \label{P_T}
\end{figure}

Agreement between our calculated Hugoniot and experimental data is
revealed in Fig. \ref{hugoniot_fig}. The region, which lies in the
density range between 1.7 and 1.8 g/cm$^{3}$, should be noticed.
This region is accompanied with a rapid increase in pressure, which
is attributed to the incipient dissociation of benzene and the
formation of diamond-like nanoparticles. The transition of atomic
structure results in the change of physical properties. Diamond is
the hardest material, and the compression in such system will
consequently lead to the rapid increase in pressure. Experiments
indicate that benzene decomposes at 13 GPa
\cite{PBX:Hauver:1965,PBX:Dick:1970,PBX:Warnes:1970,PBX:Yakusheva:1971,PBX:Holmes:1990},
while our QMD simulations provide the value of 11 GPa. We further
examine the EOS data at the densities ranging from 2.1 to 2.4
g/cm$^{3}$ as shown in Fig. \ref{P_T}. The pressure shows a
systematic behavior in terms of the density and temperature, except
for the region ($\rho$=2.3 $\sim$ 2.4 g/cm$^{3}$, $T$=3500 $\sim$
4000 K, and $P$$\approx$50 GPa). Due to the rapid C-H bond breaking
and the formation of atomic and molecular hydrogen, this region is
featured by ($\partial P/\partial T$)$_{V}<$0.

\begin{table}[!htbp]
\centering \caption{Principal Hugoniot points derived from DFT-MD
simulations at a series of density ($\rho$), pressure ($P$), and
temperature ($T$). }
\begin{tabular}{ccc}
\hline\hline
$\rho$ (g/cm$^{3}$)& $P$ (GPa) & $T$ (K)  \\
\hline
 1.2  &  4.3  &  598   \\
 1.4  &  7.9  &  814   \\
 1.6  & 10.0  &  1093  \\
 1.7  & 10.9  &  1200  \\
 1.8  & 19.3  &  1324  \\
 1.9  & 19.2  &  1437  \\
 2.1  & 33.7  &  2963  \\
 2.2  & 43.2  &  3370  \\
 2.3  & 54.4  &  3749  \\
 2.4  & 67.2  &  4334  \\
\hline\hline
\end{tabular}
\label{H_data}
\end{table}

\begin{figure}[!htbp]
\centering
\includegraphics*[width=\textwidth]{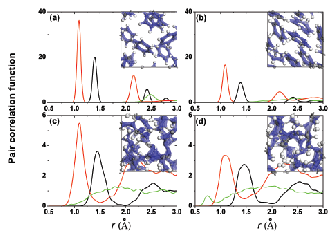}
\caption{Calculated pair correlation function for C-C (black line),
C-H (red line), H-H (green line) at four densities of benzene along
the principal Hugoniot. The atomic structure, where carbon and
hydrogen atoms are denoted by gray and white balls respectively, and
the relative iso-surface of charge density (blue regimes) are also
provided in the insets. (a) $\rho$=0.874 g/cm$^{3}$, $T$=298 K; (b)
$\rho$=1.4 g/cm$^{3}$, $T$=814 K; (c) $\rho$=2.1 g/cm$^{3}$,
$T$=2963 K; (d) $\rho$=2.3 g/cm$^{3}$, $T$=3749 K.} \label{rdf}
\end{figure}

The structural change of benzene under shock condition can be
clearly reflected by PCF, which represents the possibility of
finding a particle at a distance $r$ from a reference atom. Typical
C-C bond length in benzene molecule is 1.40 \AA, and the bond length
of C-H is 1.09 \AA\ in our calculations. In the case of diamond and
H$_{2}$, C-C and H-H have the bond length of 1.50 and 0.75 \AA,
respectively. PCF, atomic structure, and charge density distribution
at four densities of benzene along the principal Hugoniot are shown
in Fig. \ref{rdf}. At $\rho$=0.874 g/cm$^{3}$ and $T$=298 K, one can
see from Fig. \ref{rdf}(a) that the PCF is featured by two peaks at
1.09 and 1.40 \AA, which represent the typical bond length of C-H
and C-C in benzene molecule, respectively. These two peaks begin to
reduce in amplitude and get broaden due to the thermal excitation at
higher pressure [as shown in Fig. \ref{rdf}.(b)]. For the densities
$\rho<$1.7 g/cm$^{3}$ and temperatures $T<$1200 K, benzene turns out
to retain in its molecular shape. This fact could be established
through analysis of the atomic configuration and typical $\pi$ bond
suggested by charge density. When $\rho>$1.7 g/cm$^{3}$ and $T>$1200
K, diamond-like nanoparticles with vacancies and H substitutions are
formed in the system. At this stage, C-C bond length changes into
1.50 \AA\ [Fig. \ref{rdf}.(c)], which is a typical value of diamond,
and the $\pi$ bond converts into $sp$ hybrid bond. Benzene
decomposes under the present condition, and the reference pressure
is 11 GPa. With further increase in pressure, the dissociation of
C-H bond becomes remarkable, and the formation of H$_{2}$ molecule
is suggested by the peak at $r$=0.75 \AA, as shown in Fig.
\ref{rdf}(d).

\section{DYNAMIC CONDUCTIVITY AND OPTICAL PROPERTIES} \label{sec-optics}

The real part of the dynamic conductivity $\sigma_{1}(\omega)$ is
derived from the Kubo-Greenwood formula:

\begin{eqnarray} \label{real-conductivity}
\sigma_{1}(\omega)=\frac{2\pi
e^{2}\hbar^{2}}{3m^{2}\omega\Omega}&\sum\limits_{\textbf{k}}w(\textbf{k})\sum\limits_{j=1}^{N}\sum\limits_{i=1}^{N}\sum\limits_{\alpha=1}^{3}[f(\epsilon_{i},\textbf{k})-f(\epsilon_{j},\textbf{k})]
\cr
&\times|\langle\Psi_{j,\textbf{k}}|\nabla_{\alpha}|\Psi_{i,\textbf{k}}\rangle|^{2}\delta(\epsilon_{j,\textbf{k}}-\epsilon_{j,\textbf{k}}-\hbar\omega),
\end{eqnarray}
where $f(\epsilon_{i},\textbf{k})$ describes the occupation of the
$i$th band, with the corresponding energy $\epsilon_{i,\textbf{k}}$
and the wavefunction $\Psi_{i,\textbf{k}}$ at $\textbf{k}$, and
$w(\textbf{k})$ is the $\textbf{k}$-point weighting factor. The
imaginary part of the dynamic conductivity $\sigma_{2}(\omega)$
follows from the Kramer-Kr\"{o}nig relationship. Then, the real and
imaginary parts of the dielectric function are given by
$\epsilon_{1}(\omega)=1-\frac{1}{\epsilon_{0}\omega}\sigma_{2}(\omega)$
and
$\epsilon_{2}(\omega)=\frac{1}{\epsilon_{0}\omega}\sigma_{1}(\omega)$,
respectively, from which the optical constants such as the
refractive index $n(\omega)$, extinction coefficient $k(\omega)$ and
reflectivity $r(\omega)$ may now be derived ($\epsilon
_{1}$=$n^{2}\mathtt{-}k^{2}$, $\epsilon _{2}$=$2nk$, and
$r\mathtt{=}\frac{(1-n)^{2}+k^{2}}{(1+n)^{2}+k^{2}}$).

\begin{figure}[!htbp]
\centering
\includegraphics*[width=\textwidth]{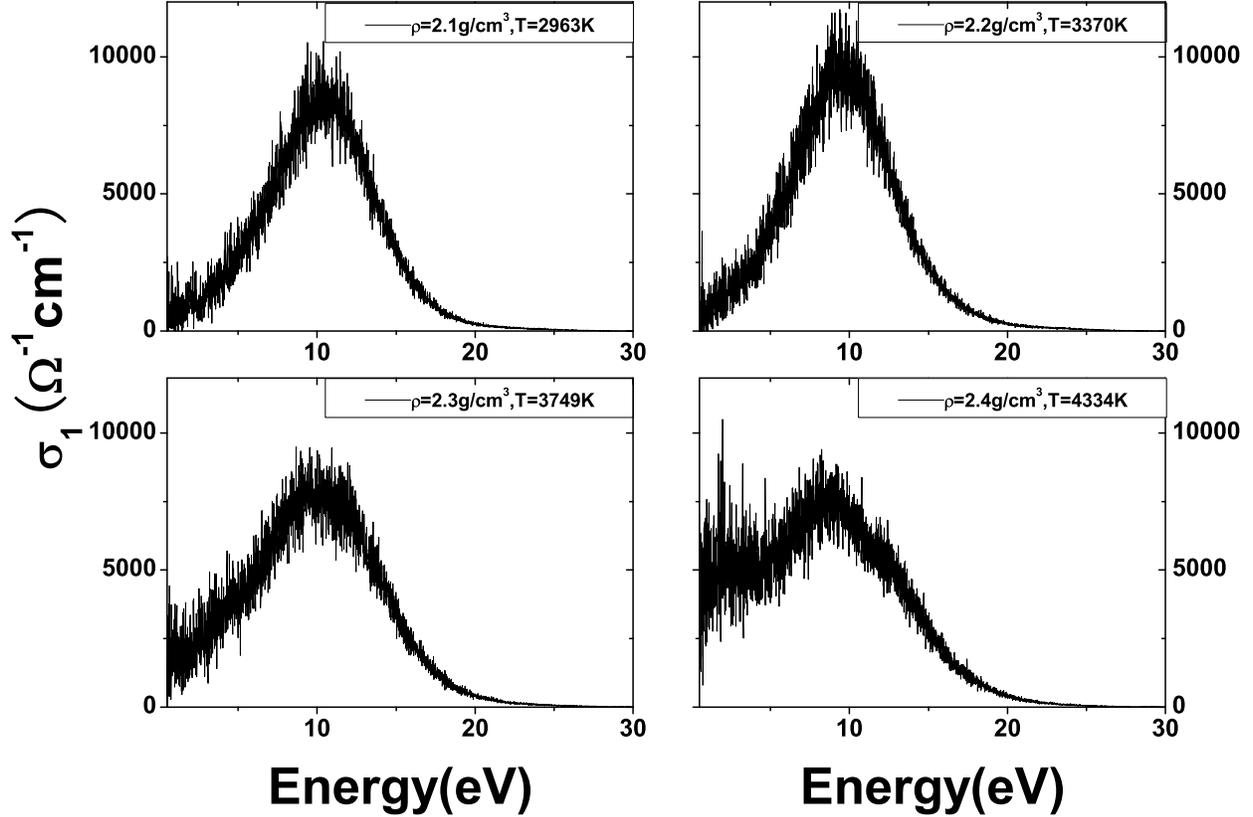}
\caption{$\sigma_{1}(\omega)$ along the principal Hugoniot. }
\label{conductivity}
\end{figure}

\begin{figure}[!htbp]
\centering
\includegraphics*[width=\textwidth]{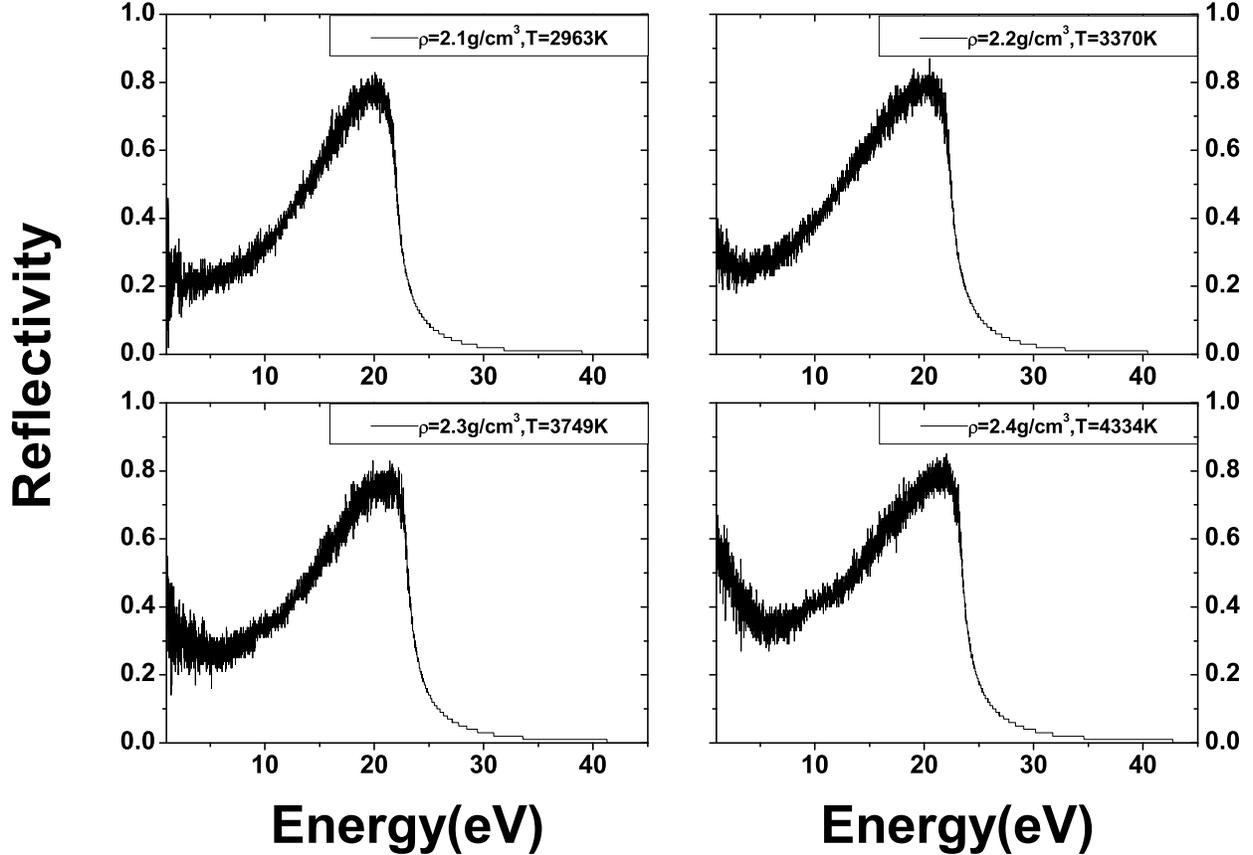}
\caption{Optical reflectivity of benzene along the principal
Hugoniot. } \label{reflectivity}
\end{figure}

The behavior of $\sigma_{1}(\omega)$ at different Hugoniot points is
summarized in Fig. \ref{conductivity}. Overall, we find that the
real part of dynamic conductivity [$\sigma_{1}(\omega)$] is featured
by a main peak around 10.0 eV and approach to vanish for the energy
higher than 25.0 eV. For the pressure lower than 30 $\sim$ 40 GPa
($\rho<$2.1 $\sim$ 2.2 g/cm$^{3}$), the dc conductivity, which could
be given as:
$\sigma_{dc}=\lim\limits_{\omega\rightarrow0}\sigma_{1}(\omega)$, is
zero, which obviously shows the insulating nature of benzene at low
pressure regime. With the increase of pressure along the principal
Hugoniot ($\rho$=2.3 $\sim$ 2.4 g/cm$^{3}$), three prominent new
features occur in the frequency-dependent conductivity: (i) The main
peak in $\sigma_{1}(\omega)$ moves towards lower energies; (ii)
There tends to develop an additional little peak around the energy
of 2.5 eV, which again signifies the occurrence of diamond-like
structure in the system; (iii) The dc conductivity is no longer
zero. This indicates that nonmetal-metal transition, which is
associated with the rapid C-H bond breaking, happens at the
high-pressure Hugoniot points around 50 GPa ($\rho$$\sim $2.3
g/cm$^{3}$ and $T\sim $3700 K), consistent with the predicted value
derived from experiment measurement \cite{PBX:Nellis:2001}.
Furthermore, the optical reflectivity of benzene along the principal
Hugoniot is shown in Fig. \ref{reflectivity}. The pressure-induced
change in reflectivity can be clearly indicated in the energy range
of 1.0$\sim$5.0 eV (the reference wave length is correspondingly
from 250 to 1250 nm). For the photon energy $E$=1.0 eV, a
considerable increase in reflectivity (from 0.2 to 0.6), which is
attributed to high pressure induced nonmetal-metal transition, is
observed along the principal Hugoniot.

\section{CONCLUSION} \label{sec-con}

In summary, the QMD simulations presented here demonstrate the
properties of shock compressed benzene. The EOS and principal
Hugoniot are determined according to the simulated results, and
the results show excellent agreement with the available
experiments. Our calculations have shown that the benzene molecule
begins to dissociate around 11 GPa. The considerable C-H bond
breaking appears at the pressure around 50 GPa. The nonmetal-metal
transition also occurs at this stage, which will lead to the
increase in optical reflectivity from 0.2 to 0.6.

\begin{acknowledgments}
 This work was supported by the Foundation for Development of
Science and Technology of China Academy of Engineering Physics
under Grant No. 2009B0301037.
\end{acknowledgments}


\begin{thebibliography}{10}

\bibitem{PBX:Elert:2003}M. L. Elert, S. V. Zybin, and C. T. White, J. Chem. Phys.,
\textbf{118}, 9795 (2003).

\bibitem{PBX:Hubbard:1980}W. B. Hubbard, Science,
\textbf{214}, 145 (1980).

\bibitem{PBX:Stevenson:1982}D. J. Stevenson, Annu. Rev. Earth Planet Sci.,
\textbf{14}, 257 (1982).


\bibitem{PBX:Mimura:1994}K. Mimura, M. Kato, and R. Sugisaki, Geophys. Res.
Lett.,
\textbf{21}, 2071 (1994).


\bibitem{PBX:Mimura:1995a}
K. Mimura, Geochim. Cosmochim. Acta, \textbf{59}, 579 (1995).

\bibitem{PBX:Dick:1969}
R. Dick, J. Chem. Phys., \textbf{52}, 6021 (1969).

\bibitem{PBX:Hauver:1965}
G. E. Hauver, J. Appl. Phys., \textbf{36}, 2113 (1965).

\bibitem{PBX:Dick:1970}
R. D. Dick, J. Chem. Phys., \textbf{52}, 6021 (1970).

\bibitem{PBX:Warnes:1970}
R. H. Warnes, J. Chem. Phys., \textbf{53} 1088 (1970).

\bibitem{PBX:Yakusheva:1971}
O. B. Yakusheva, V. V. Yakushev, and A. N. Dremin, High Temp.-High
Press., \textbf{3} 261 (1971).

\bibitem{PBX:Holmes:1990}
N. C. Holmes, G. Otani, P. McCandless, and F. H. Ree, High Press.
Res., \textbf{5}, 669 (1990).

\bibitem{PBX:Nellis:1984}
W. J. Nellis, F. H. Ree, R. J. Trainor, A. C. Mitchell, and M. B.
Boslough, J. Chem. Phys., \textbf{80} 2789 (1984).

\bibitem{PBX:Mimura:1995b}
K. Mimura, M. Kato, and R. Sugisaki, Earth Planet. Sci. Lett.,
\textbf{133} 265 (1995).

\bibitem{PBX:Davis:1996}
L. L. Davis and K. R. Brower, J. Phys. Chem., \textbf{100} 18775
(1996).

\bibitem{PBX:Nellis:2001}
W. J. Nellis, D. C. Hamilton, and A. C. Mitchell, J. Phys. Chem.,
\textbf{115} 1015 (2001).

\bibitem{PBX:Bickham:2000}
S. R. Bickham, J. D. Kress, and L. A. Collins, J. Phys. Chem.,
\textbf{112} 9695 (2000).

\bibitem{PBX:Maillet:2008}
J. B. Maillet and N. Pineau, J. Phys. Chem., \textbf{128} 224502
(2008).

\bibitem{PBX:Mattsson:2006}
T. R. Mattsson and M. P. Desjarlais, Phys. Rev. Lett., \textbf{97}
017801 (2006).

\bibitem{PBX:Kietzmann:2007}
A. Kietzmann, B. Holst, R. Redmer, M. P. Desjarlais, and T. R.
Mattsson, Phys. Rev. Lett., \textbf{98} 190602 (2007).

\bibitem{PBX:Kresse:1993}
G. Kresse and J. Hafner, Phys. Rev. B, \textbf{47} R558 (1993).

\bibitem{PBX:Kresse:1996}
G. Kresse and J. Furthm{\"u}ller, Phys. Rev. B, \textbf{54} 11 169
(1996).



\bibitem{PBX:Lenosky:2000}
T. Lenosky, S. Bickham, J. Kress, and L. Collins, Phys. Rev. B,
\textbf{61} 1 (2000).

\bibitem{PBX:Bagnier:2001}
S. Bagnier, P. Blottiau, and J. Clerouin, Phys. Rev. E,
\textbf{63} 015 301(R) (2001).



\bibitem{PBX:Perdew:1991}
J. P. Perdew, \textit{Electronic Structure of Solids} (Akademie
Verlag, Berlin, 1991).

\bibitem{PBX:Blochl:1994}
P. E. Bl{\"o}chl, Phys. Rev. B, \textbf{50} 17953 (1994).

\bibitem{PBX:Kresse:1999}
G. Kresse and D. Joubert, Phys. Rev. B, \textbf{59} 1758 (1999).

\bibitem{PBX:Monkhorst:1976}
H. J. Monkhorst and J. D. Pack, Phys. Rev. B, \textbf{13} 5188
(1976).

\bibitem{PBX:Zeldovich:1966}
Y. Zeldovich and Y. Raizer, \textit{Physics of Shock Wave and High
Temperature Hydrodynamic Phenomena} (Academic Press, New York,
1966).



\end{thebibliography}

\end{document}